# YUHENG-OS: A Cloud-Native Space Cluster Operating System

Jin Zhang, Tsinghua University

Jiachen Sun, Hong Kong University of Science and Technology

Kai Liu, Tsinghua University

Linling Kuang, Tsinghua University

Jianhua Lu, Tsinghua University

*As industry and academia continue to advance spaceborne computing and communication capabilities, the formation of cloud-native space clusters (CNSCs) has become an increasingly evident trend. This evolution progressively exposes the resource management challenges associated with coordinating fragmented and heterogeneous onboard resources while supporting large-scale and diverse space applications. However, directly transplanting mature terrestrial cloud-native cluster operating system paradigms into space is ineffective due to the fragmentation of spaceborne computing resources and satellite mobility, which collectively impose substantial challenges on resource awareness and orchestration. This article presents YUHENG-OS, a cloud-native space cluster operating system tailored for CNSCs. YUHENG-OS provides unified abstraction, awareness, and orchestration of heterogeneous spaceborne infrastructure, enabling cluster-wide task deployment and scheduling across distributed satellites. We introduce a four-layer system architecture and three key enabling technologies: modeling of heterogeneous resource demands for space tasks, fragmented heterogeneous resource awareness under network constraints, and matching of differentiated tasks with multidimensional heterogeneous resources under temporal dependency constraints. Evaluation results show that, compared with representative terrestrial cloud-native cluster operating systems exemplified by Kubernetes, YUHENG-OS achieves a substantially higher task completion ratio, with improvements of up to 98%. This advantage is primarily attributed to its ability to reduce resource awareness delay by 71%.*

## Introduction

In ancient Chinese cosmology, YUHENG evokes an instrument of steadiness amid motion—an anchor for ordering the sky when the stars do not stand still. In The Canon of Shun, this idea is expressed in the phrase "*observing the heavens with XUANJI and YUHENG to align the governance of all celestial affairs.*" It reflects a simple yet powerful principle: when the environment is seemingly chaotic and inherently dynamic, effective governance depends on accurate awareness and coordinated management.

Ongoing advances in spaceborne computing and communication technologies are driving satellite systems to evolve from single-satellite, mission-specific platforms toward constellation-scale, networked infrastructures, enabling emerging applications such as spaceborne data inference and training, as well as Direct Handset to Satellite (DHTS) [1], [2]. With the continuous expansion of constellation size and the increasing diversity of onboard payload capabilities, multiple satellites in orbit are progressively forming a shared heterogeneous resource pool encompassing computation (CPU/GPU/AI accelerators), storage, networking (inter-satellite and satellite–ground laser/microwave links), etc. [3]. These resources can be sensed, abstracted, and orchestrated in a unified manner. In parallel, onboard applications are transitioning from tightly coupled, mission-specific programs to portable, containerized application units, enabling workloads such as in-orbit data training and spaceborne image processing to be deployed and migrated onboard in a lightweight fashion [1]. This evolution has given rise to the early form of cloud-native space clusters (CNSCs).

Recent momentum in both industry and academia further indicates that the construction of CNSCs has become practically feasible. From an industrial perspective, in 2025, Planet partnered with NVIDIA to integrate GPUs into its Owl remote-sensing mega-constellation, enabling real-time, in-orbit processing of high-resolution Earth observation data [4]. Starcloud successfully launched its test satellite Starcloud-1, equipped with an NVIDIA H100 GPU, into Low Earth Orbit (LEO), marking a milestone as the first attempt to conduct language model training in space using data center–class GPUs [5]. In parallel, Google announced its Project Suncatcher initiative, proposing an in-orbit computing constellation composed of 81 satellites equipped with TPUs [6], while Starlink has explicitly stated plans to integrate onboard computing payloads into its V3 satellites [7]. In China, the Chenguang-1 satellite was launched in 2025, with an onboard computing capability comparable to that of a ground-based server. Subsequent plans aim to build a space data center with an overall computing capacity exceeding 400,000 Peta Floating Point Operations Per Second (PFLOPS) [8]. In academia,

Tsinghua University has proposed the deployment of the Tsinghua Space Network (TSN) in Medium Earth Orbit (MEO), leveraging its wide-area coverage and stable communication capacity to provide sustained and reliable management and artificial intelligence services for massive volumes of spaceborne data and heterogeneous resources. The first satellite of TSN, TN-1A, was successfully launched in 2024 and is currently undergoing in-orbit validation of key enabling technologies [9].

Against this backdrop, there is a growing need for a stable resource management platform capable of bridging large-scale, diverse space applications with fragmented, heterogeneous on-orbit resources. A useful point of reference can be found in terrestrial cloud platforms, where cloud-native cluster operating systems already play a comparable role by matching massive, multi-tenant application demands with large pools of computing and storage infrastructure. The cloud-native software stack represented by Kubernetes—proposed and open-sourced by Google in 2014—has gradually become an industry standard, enabling mainstream cloud service providers to support large-scale automated deployment, elastic scaling, and operational management of containerized applications in data center environments [10]. Building on Kubernetes, Amazon offers Elastic Kubernetes Service (EKS), which is deeply integrated with its cloud infrastructure for resource orchestration and operational management [11]. Similarly, major cloud vendors, including Tencent Cloud [12] and Baidu AI Cloud [13], have adopted Kubernetes as the core foundation for building their own cluster operating systems targeting large-scale data center deployments. Huawei Cloud further extends Kubernetes through the KubeEdge framework, enabling cluster operating system capabilities to expand beyond centralized data centers to distributed nodes and heterogeneous device environments [14]. These successful precedents naturally motivate efforts to extend the cloud-native cluster operating system paradigm to CNSCs. However, owing to fundamental differences in environmental conditions and system constraints, directly porting terrestrial cloud-native cluster operating system paradigms to space is often ineffective.

One fundamental challenge arises from the fragmented nature of spaceborne computing resources. Unlike terrestrial clusters, where computing capacity is typically aggregated into a small number of large, homogeneous server clusters interconnected by high-bandwidth links, satellite computing resources are inherently distributed across a large population of spatially separated nodes. For example, in China, over 8.3 million standard server racks are deployed nationwide, collectively providing computing capacity on the order of 246 EFLOPS, yet such massive capacity is concentrated in only 10 national data center clusters [15]. In contrast, computing resources in CNSCs are spread across multiple orbital shells and thousands of satellites, each offering limited onboard capability—currently on the order of hundreds of TOPS—under strict power and thermal constraints. As the network scale increases, resource fragmentation becomes more pronounced, substantially increasing the difficulty of resource management, particularly for the Network Operation and Control Centers (NOCCs) to maintain timely and accurate visibility into the resource states of CNSCs.

Another critical challenge stems from satellite mobility, which results in constrained visibility windows and intermittent connectivity. For low Earth orbit systems, satellite–ground communications typically last only several minutes to on the order of ten minutes per pass, making it impractical for satellite–ground state synchronization to rely on high-frequency, stable heartbeat mechanisms in the same manner as terrestrial cloud-native clusters. Inter-satellite links exhibit similarly transient availability, as visibility windows may range from minutes to hours depending on orbital configurations, while link quality continuously varies with relative geometry, thereby preventing the establishment of stable, fixed high-throughput connections comparable to data-center optical interconnects. These mobility-driven network dynamics fundamentally challenge terrestrial cloud-native cluster operating systems, which are largely centered on computation and storage with relatively limited modeling of communication constraints. Consequently, in CNSCs, communication resources must be explicitly modeled and orchestrated as constraints that are as critical as computation and storage, and, in certain scenarios, may even become the dominant limiting factor.

In summary, the fragmentation and mobility of spaceborne resources constitute a central challenge for CNSCs, which involves not only obtaining timely and accurate awareness of fragmented resource states, but also efficiently orchestrating heterogeneous resources and diverse tasks under highly dynamic conditions. Remarkably, these two aspects align closely with the principles embodied by XUANJI and YUHENG. Accordingly, we propose an efficient resource management platform for CNSCs, named YUHENG-OS, which serves as a key enabling layer between space applications and space cluster infrastructures under network constraints by providing unified abstraction, awareness, and orchestration of heterogeneous resources, as well as supporting cluster-wide deployment and scheduling of space tasks. The main contributions of this article are threefold:

(1) We propose a four-layer architecture for YUHENG-OS tailored to CNSCs, which effectively bridges the space application layer and the underlying space cluster infrastructure by providing unified support for cluster scalability, awareness of fragmented resources, heterogeneous resource orchestration, and task demand analysis.

(2) We develop three key technologies, including modeling of heterogeneous resource demands for space tasks, fragmented heterogeneous resource awareness under network constraints, and matching of differentiated tasks with multidimensional heterogeneous resources under temporal dependency constraints.

(3) We conduct comprehensive evaluations to assess the effectiveness of YUHENG-OS. The results demonstrate that, compared with terrestrial cloud-native cluster operating systems exemplified by Kubernetes, YUHENG-OS achieves significant improvements in task completion ratio.

The remainder of this article is organized as follows. Section II introduces the proposed YUHENG-OS architecture. Section III presents the three key technologies. Section IV reports and analyzes the simulation results. Section V concludes the article.

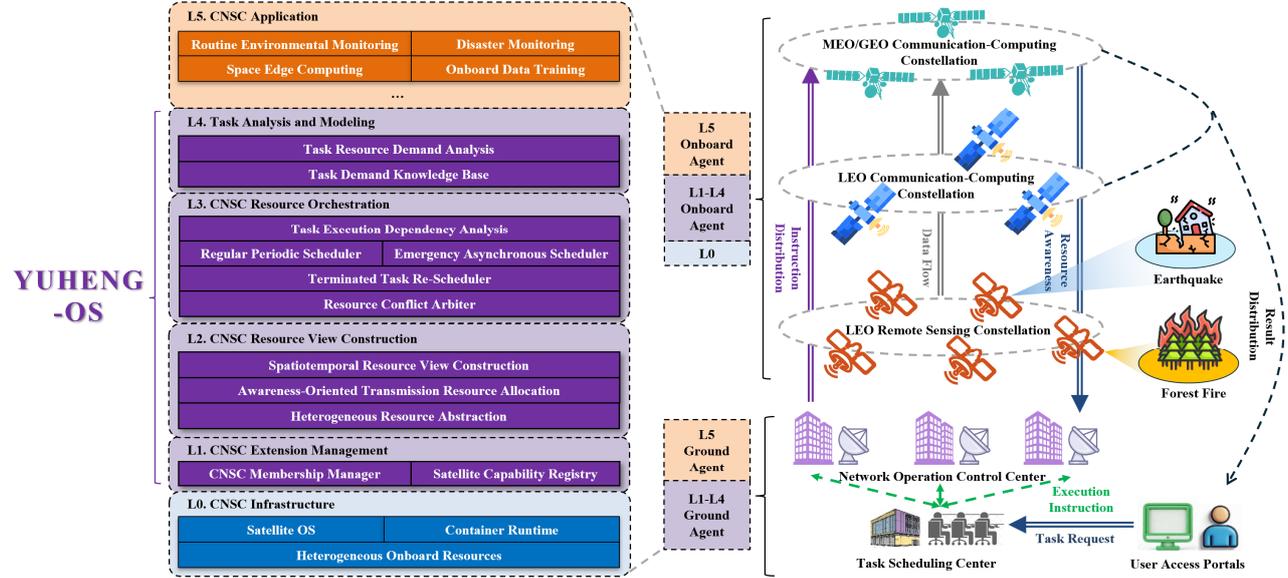

**Figure 1** *Architecture overview of YUHENG-OS and its deployment in a CNSC.*

## YUHENG-OS Architecture Overview

A CNSC system comprises satellites distributed across multiple orbital regimes, including LEO, MEO, and Geostationary Earth Orbit (GEO), and spanning diverse functional roles such as remote sensing–computing and communication–computing. Its ground segment includes an NOCC for each constellation, a unified Task Scheduling Center, and multiple User Access Portals.

Within CNSC, the YUHENG-OS serves as the core resource management layer, bridging the underlying infrastructure layer (L0) and the application layer (L5), as shown in Figure 1. It is designed to provide unified abstraction, awareness, and orchestration of fragmented, heterogeneous spaceborne resources, thereby enabling on-demand service provisioning for large-scale space applications.

From a system perspective, L0 resides beneath YUHENG-OS and constitutes the foundational infrastructure for resource provisioning and task execution environments. It includes heterogeneous onboard resources, the satellite OS, and the container runtime, jointly formed by per-satellite heterogeneous spaceborne resources, the satellite operating system, and the container runtime. Specifically, heterogeneous onboard resources comprise spaceborne computing units (e.g., CPUs and GPUs), storage resources, inter-satellite and satellite–ground laser/microwave communication links, as well as mission-specific payloads such as remote sensing sensors. The satellite OS manages local resources and reports resource information to YUHENG-OS, while the container runtime provides a unified execution environment for cloud-native applications deployed onboard. Together, these components form the resource and runtime basis for YUHENG-OS.

Above YUHENG-OS, L5 represents the application layer that supplies task inputs and service requests, encompassing a broad spectrum of cloud-native space applications, including but not limited to routine environmental monitoring, disaster monitoring, space edge computing, and onboard data training. In addition, an L5 Onboard Agent is deployed on each satellite to receive task execution directives and manage the task execution lifecycle, while an L5 Ground Agent is deployed at each NOCC and the Task Scheduling Center to accept user task requests and distribute task execution instructions to corresponding satellites.

Functionally, YUHENG-OS is organized into four logical layers: CNSC Extension Management (L1), CNSC Resource View Construction (L2), CNSC Resource Orchestration (L3), and Task Analysis and Modeling (L4), which are deployed in the form of onboard agents and ground agents. The following subsections describe the design rationale and core functionalities of these four layers in detail, followed by an overview of the deployment architecture and operational workflow of YUHENG-OS within the overall system.

### L1. CNSC Extension Management

The CNSC Extension Management layer provides controlled cluster expansion and capability registration for

CNSC. It consists of a CNSC Membership Manager and a Satellite Capability Registry, which together enable heterogeneous satellites to elastically join and leave the cluster. The CNSC Membership Manager maintains node lifecycle states and cluster membership consistency under intermittent connectivity, while the Satellite Capability Registry maintains abstract descriptions of node capabilities across computing, storage, communication, etc. These capability descriptions reflect configured capacity bounds rather than instantaneous resource availability, providing a stable reference for subsequent resource abstraction and orchestration. This layer exposes a unified, software-defined interface for scalable cluster extension in YUHENG-OS. It is implemented via both onboard and ground agents, where the onboard agent executes local join/leave handshakes, maintains per-satellite lifecycle state, and reports capability descriptors, while the ground agent performs cluster-wide membership coordination, admission control, and global capability registration for management and control-plane consistency.

*L2. CNSC Resource View Construction*

The CNSC Resource View Construction layer elevates fragmented, heterogeneous resources into a cluster-level resource view suitable for scheduling and orchestration. Heterogeneous Resource Abstraction module first provides a unified abstraction and virtualization of all heterogeneous resources across cluster nodes, producing per-satellite resource profiles that standardize computing, storage, communication, and sensing capabilities. Building upon these per-satellite profiles, Awareness-Oriented Transmission Resource Allocation module assigns appropriate transmission channels for each satellite's resource information and delivers the resource reports to the NOCC. Finally, Spatiotemporal Resource View Construction module consolidates the received multi-satellite resource reports into an integrated spatiotemporal resource view, passed to the L3 layer as real-time resource-state input to support resource-aware orchestration. It is implemented via both onboard and ground agents, where the onboard agent performs local heterogeneous resource abstraction, packages and periodically publishes per-satellite resource profiles, and selects uplink/downlink telemetry channels, while the ground agent receives multi-satellite reports at the NOCC, normalizes and aggregates them, and constructs the cluster-level spatiotemporal resource view for upstream orchestration.

*L3. CNSC Resource Orchestration*

The CNSC Resource Orchestration layer matches task requirements with suitable resources and generates executable scheduling plans under network-constrained conditions and task temporal dependency constraints. Task Execution Dependency Analysis module first analyzes the temporal dependencies across different execution stages of a task and derives the resource demands of each stage across heterogeneous resources. Based on this analysis, the Regular Periodic Scheduler performs periodic scheduling for routine tasks, while the Emergency Asynchronous Scheduler enables event-triggered asynchronous scheduling for urgent tasks to accelerate response time. To address scheduling failures and execution interruptions caused by emergency task preemption or resource faults, the Terminated Task Re-Scheduler re-plans terminated tasks to restore execution continuity. Meanwhile, the Resource Conflict Arbiter resolves resource contention arising during task scheduling, ensuring consistent and feasible orchestration decisions across competing tasks. It is implemented via both onboard and ground agents, where the ground agent conducts cluster-level planning and decision-making (including periodic and event-triggered scheduling, conflict resolution, and re-scheduling) using the global resource view, while the onboard agent enforces allocated plans locally by translating schedules into executable actions, monitoring execution status, and feeding back preemption, failures, and realized execution traces for closed-loop orchestration.

*L4. Task Analysis and Modeling*

In contrast to terrestrial cloud-native systems, space-based clusters operate under tightly constrained resource budgets, making over-provisioning neither feasible nor efficient. Consequently, resource management must move beyond upper-bound allocation and instead leverage an explicit understanding of how resource provisioning levels affect task completion quality (e.g., completion latency and model accuracy), which motivates the Task Analysis and Modeling layer. This layer provides task-level intelligence by modeling resource consumption profiles under different task completion qualities, thereby enabling fine-grained resource orchestration from the demand side. The Task Resource Demand Analysis module receives task requests from the L5 layer and, based on the Task Demand Knowledge Base, characterizes each task by identifying required resource types and estimating the corresponding resource consumption needed to satisfy specified task completion quality requirements across computing, storage, and communication dimensions. The Task Demand Knowledge Base stores empirical mappings between target task quality levels and the associated heterogeneous resource consumption, serving as a calibrated reference for orchestration. It is implemented via both onboard and ground agents, where the ground agent maintains and updates the Task Demand Knowledge Base using system-wide execution feedback and provides quality-constrained resource demand estimates to the L3 layer, while the onboard agent records task-level runtime feedback—such as realized resource consumption and achieved completion quality—and reports them to continuously refine demand estimates under given quality targets.

## Key Technologies of YUHENG-OS

In this section, we discuss a set of key enabling technologies of YUHENG-OS that address the fundamental challenges of operating CNSCs.

### Modeling of Heterogeneous Resource Demands for Space Tasks

In space environments, computing, storage, and communication resources are persistently constrained. Directly adopting the coarse-grained reservation strategy commonly used in terrestrial cluster operating systems—where resources are provisioned according to worst-case demand bounds—often leads to prolonged resource idleness, thereby reducing overall system throughput and task completion efficiency. This motivates the need for fine-grained modeling of multi-dimensional resource demands over the full task lifecycle, explicitly characterizing how different task types consume resources during sensing, processing, and data transmission or downlink stages. Such modeling provides an actionable foundation for efficient scheduling and orchestration under tight resource constraints.

To this end, we consider constructing a model- and data-driven task resource demand knowledge base for space workloads, as shown in Figure 2. For representative space processing tasks, the knowledge base employs systematic task profiling and parameter exploration to cover, as comprehensively as possible, feasible resource provisioning configurations. Through repeated evaluations, it assesses key performance indicators such as timeliness and accuracy, capturing both expected performance and variability, and derives stage-aware resource demand curves. Meanwhile, observations from in-orbit execution and task logs are continuously incorporated to refine and calibrate these models, allowing the knowledge base to progressively converge toward real-world resource consumption characteristics. With this capability, newly submitted tasks can query their lifecycle-wide, multi-dimensional resource requirements prior to execution, enabling demand-driven provisioning and fine-grained orchestration that substantially reduces resource idle periods and improves overall resource utilization.

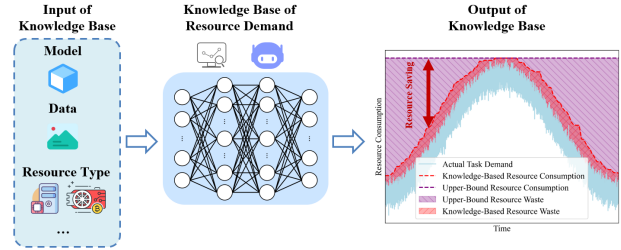

**Figure 2** *Knowledge-base–driven task demand modeling and its impact on resource allocation efficiency in space clusters.*

### Fragmented Heterogeneous Resource Awareness under Network Constraints

To support task planning and resource orchestration, resource states of satellites in the cluster must be aggregated at the NOCC with controllable latency and sufficient accuracy. However, cloud-native space clusters are characterized by wide-area node distribution, low-rate and intermittently available links, and highly dynamic and fragmented resources. Under these conditions, the uniform periodic heartbeat mechanisms commonly adopted by terrestrial cloud-native cluster operating systems—which rely on persistent and high-bandwidth control–data plane connectivity—become ineffective and poorly scalable. As a result, resource awareness and aggregation mechanisms tailored to network-constrained, time-varying environments emerge as a key enabling technology.

To address these constraints, we adopt a multi-domain awareness architecture anchored by MEO/GEO satellites, leveraging their wide coverage and relatively stable topology to form reliable transmission channels, as shown in Figure 3. Domain partitioning and inter-domain coordination explicitly account for link conditions and control-plane load. On this basis, differentiated resource state reporting strategies are employed, in which reporting granularity and frequency are adaptively determined according to resource type and volatility. Specifically, resource profiles and change models are constructed for each node–resource pair: rapidly varying resources are monitored with shorter sensing intervals, while slowly changing or long-term stable resources are reported at longer intervals or via event-driven updates. This approach significantly reduces communication overhead while preserving high-fidelity awareness of critical resource states, providing reliable inputs for subsequent scheduling decisions.

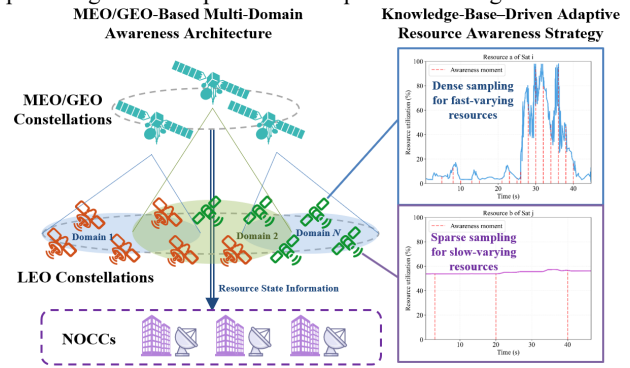

**Figure 3** *MEO/GEO-Based Multi-Domain Awareness Architecture and Knowledge-Base-Driven Adaptive Resource Awareness Strategy.*

### Matching of Differentiated Tasks with Multidimensional Heterogeneous Resources under Temporal Dependency Constraints

Matching space tasks to resources is inherently a multi-dimensional knapsack problem, where heterogeneous resources,

spanning computing, storage, communication, and sensing, need to be allocated to multi-stage task pipelines under temporal dependency constraints. Terrestrial cloud-native cluster operating systems typically operate over stable, high-bandwidth interconnects and relatively simple execution stages, and therefore adopt task-triggered, compute–storage-oriented orchestration. These conditions do not hold for space clusters, where link availability is windowed and time-varying, topology evolves continuously, and task stages exhibit distinct resource preferences. This motivates a differentiated orchestration strategy that explicitly accounts for temporal constraints and multi-resource coupling.

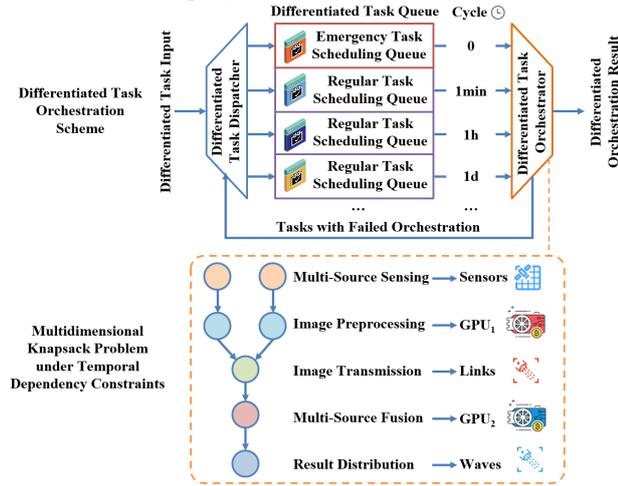

**Figure 4** *Differentiated Task Orchestration Scheme and Multidimensional Knapsack Problem under Temporal Dependency Constraints with a Multi-Source Remote Sensing Fusion Task as an Example.*

We adopt a hierarchical orchestration framework that combines synchronous pre-planning for regular tasks with asynchronous fast reaction for urgent tasks, as shown in Figure 4. Predictable, recurring workloads are handled through periodic planning and rolling updates, whereas time-critical tasks trigger rapid reactive scheduling to minimize response latency. To balance response speed and orchestration complexity, regular tasks further employ tiered planning cycles: high-priority tasks are replanned more frequently, while lower-priority tasks use longer cycles to reduce control and computation costs. Internally, each task is modeled as a directed acyclic graph (DAG), where nodes represent processing stages and edges capture dependency constraints. The processing stages are subject to strict temporal ordering constraints, such that later stages cannot commence until prerequisite stages are completed. For example, a multi-source remote-sensing fusion task can be decomposed into a sequence of ordered execution stages, including multi-source sensing, image preprocessing, image transmission, multi-source fusion, and result distribution, each exhibiting different demands and affinities for heterogeneous resources. Orchestration decisions must therefore respect temporal dependency constraints, which substantially increases decision complexity. In practice, scalable online operation can be supported using low-complexity approximation approaches, such as distributed heuristics or learning-driven policies. Moreover, due to the heterogeneity of onboard computing resources, different tasks exhibit distinct preferences for specific computing devices and achieve varying execution efficiencies. To fully exploit the capabilities of heterogeneous devices, heterogeneity needs to be explicitly accounted for during orchestration.

## Results and Discussions

To quantitatively evaluate the performance of YUHENG-OS in spaceborne cloud-native environments, we compare it with Kubernetes, a representative terrestrial cluster operating system. A dedicated simulation framework is constructed to model a CNSC. As summarized in Table 1, network size from 600 to 6,000 satellites is considered across LEO, MEO, and GEO orbits. LEO and MEO adopt the Starlink and TSN orbital configurations, respectively, while GEO satellites are uniformly distributed along the equatorial plane. The total computing capacity of the CNSC is set to 3000 GB/s and distributed across all satellites, resulting in increasing computing dispersion as the network size grows. Rather than characterizing computational capability using conventional metrics such as Operations Per Second (OPS) or FLOPS, we express it in terms of the equivalent data processing throughput per second. Inter-satellite links are configured with microwave (100–500 kbps) and laser (5–20 Gbps) capacities, and satellite-to-ground links are set to 1 Gbps. Task arrivals range from 500 to 4,000, with four priority levels, where level 4 denotes emergency tasks. Each task represents a typical multi-source remote-sensing image fusion workload, generating 5 GB of raw data that is reduced to 20 MB after onboard processing. Performance is evaluated using the weighted task completion ratio, which captures the system's ability to guarantee prioritized tasks while efficiently utilizing heterogeneous computing and communication resources.

| Parameter | Value |
|---|---|
| Total Number of Satellites | From 600 to 6,000 (including LEO, MEO, and GEO) |
| Total Computing Capacity | 3,000 GB/s (distributed across all satellites) |
| Inter-Satellite Link Capacity | {100, 200, 500} kbps (microwave) {5, 10, 20} Gbps (laser) |
| Satellite-to-Ground Link Capacity | 1 Gbps |
| Number of Tasks | From 500 to 4,000 |
| Task Priority Levels | Regular Task: 1, 2, 3, Emergency Task: 4 |
| Origin Data Volume | 5 GB |
| Processed Data Volume | 20 MB |

**Table 1** *Parameter Settings.*

Figure 5 shows that, as the network size increases, YUHENG-OS and Kubernetes exhibit markedly different trends in weighted task completion ratio. With increasing network size, the weighted task completion ratio of YUHENG-OS consistently improves, achieving a maximum gain of 67%, whereas that of Kubernetes degrades significantly, with a reduction of approximately 50%. With a network size of 6000 and 4000 arrival tasks, the performance gap between YUHENG-OS and Kubernetes widens to 98%.

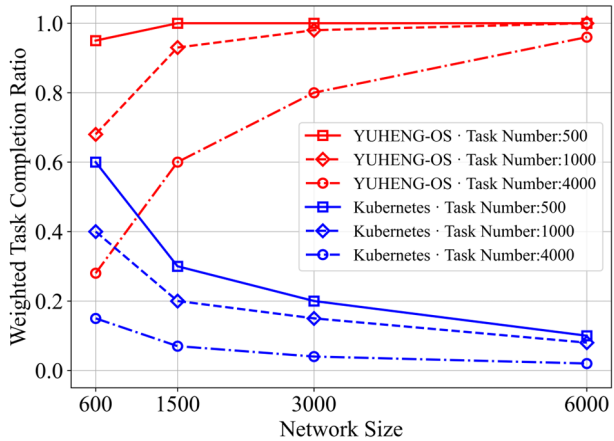

**Figure 5** *Weighted Task Completion Ratio with Varying Network Size and Task Number.*

To further identify the root causes of this performance disparity, we conduct a comparative analysis of the average resource awareness delay and the awareness-induced scheduling failure ratio for YUHENG-OS and Kubernetes. The average resource awareness delay characterizes the latency incurred by the NOCCs in acquiring a global resource view across the constellations. The awareness-induced scheduling failure ratio is defined as the proportion of scheduling failures attributable to awareness delays relative to the total number of failed tasks, thereby directly quantifying the impact of awareness latency on scheduling outcomes. As illustrated in Figure 6, the average resource awareness delay of Kubernetes increases sharply with network size. When the network size reaches 6000, the resource view obtained by the NOCCs lags the actual onboard resource states by an average of 48s, resulting in an awareness-induced relative scheduling failure ratio as high as 82%. In contrast, YUHENG-OS demonstrates strong robustness to network scaling. Under the same 6000-satellite configuration, its average resource awareness delay is only 5 s, substantially lower than that of Kubernetes, and the awareness-induced relative scheduling failure ratio is reduced by approximately 71%.

These results indicate that the improvement in task completion ratio achieved by YUHENG-OS over Kubernetes is primarily attributed to its significantly reduced awareness delay, underscoring that timely and accurate resource awareness is critical to the effective management of CNSCs. This advantage stems from the multi-domain adaptive periodic awareness mechanism enabled by MEO/MEO constellations. Furthermore, by incorporating fine-grained task modeling and explicitly accounting for inter-stage temporal dependencies and the dynamic satellite network topology during resource orchestration, YUHENG-OS enables precise resource allocation at each execution stage, thereby sustaining high task completion ratios under large-scale and high-load conditions.

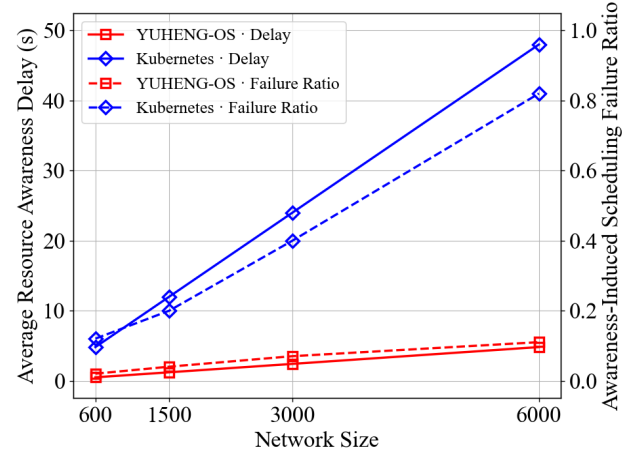

**Figure 6** *Average Resource Awareness Latency and Awareness-Induced Scheduling Failure Ratio with Varying Network Size (Task Number = 4000).*

## Conclusion

This article investigated the emerging paradigm of cloud-native space clusters and examined the limitations of directly applying terrestrial cluster operating systems to space environments, with fragmented spaceborne computing resources and satellite mobility. To address these challenges, we proposed YUHENG-OS, a cloud-native space cluster operating system that establishes a unified resource management pathway spanning from space applications down to the underlying space cluster infrastructures through a four-layer architecture comprising CNSC Extension Management, Resource View Construction, Resource Orchestration, and Task Analysis and Modeling. Building on this architecture, we further introduced three key enabling technologies: modeling of heterogeneous resource demands for space tasks, fragmented heterogeneous resource awareness under network constraints, and matching of differentiated tasks with multidimensional heterogeneous resources under temporal dependency constraints. Simulation results demonstrate that YUHENG-OS significantly outperforms representative terrestrial solutions exemplified by Kubernetes in terms of task completion ratio, primarily due to its substantially lower resource awareness latency.

These findings highlight the need for CNSC-oriented operating systems to explicitly incorporate network constraints, temporal dependencies, and resource heterogeneity into their core design. In this article, YUHENG-OS provides an OS-level foundation that delivers stable and unified resource management for inherently fragmented and dynamic space cluster infrastructures, thereby enabling scalable and efficient cloud-native space clusters.

## Acknowledgements

This work was supported in part by the National Natural Science Foundation of China (Grant No. 62341130), in part by the Tsinghua University Initiative Scientific Research Program, and in part by the Shanghai Municipal Science and Technology Major Project.

## Author Information

*Jin Zhang* (jin-zhan22@mails.tsinghua.edu.cn) is currently working toward the Ph.D. degree with the Department of Electronic Engineering, Tsinghua University, Beijing, China. He received the B.S. degree from Beijing University of Posts and Telecommunications, Beijing, China, in 2022. His current research interest focuses on space-based computing networks.

*Jiachen Sun* (sjc20@tsinghua.org.cn) is currently a postdoctoral researcher with The Hong Kong University of Science and Technology, Hong Kong, China. He received the Ph.D. degree from the Department of Electronic Engineering, Tsinghua University, Beijing, China, in 2026, and the B.S. degree from Xidian University, Xi'an, China. His current research interest is space-based computing networks.

*Kai Liu* (liukaiv@mail.tsinghua.edu.cn) is currently an Associate Research Fellow with the Beijing National Research Center for Information Science and Technology, Tsinghua University, Beijing, China. He received the B.S. and M.S. degrees from Xidian University, Xi'an, China, in 2009 and 2012, respectively, and the Ph.D. degree from Tsinghua University, Beijing, China, in 2016. His current research interests include space information networks and on-board switching.

*Linling Kuang* (kll@mail.tsinghua.edu.cn) is currently a Research Fellow with the Beijing National Research Center for Information Science and Technology, Tsinghua University, Beijing, China. She received the B.S. and M.S. degrees from the National University of Defense Technology, Changsha, China, in 1995 and 1998, respectively, and the Ph.D. degree from Tsinghua University, Beijing, China, in 2004. Her research interest is satellite communications.

*Jianhua Lu* (lhh-dee@mail.tsinghua.edu.cn) is currently a Professor with the Department of Electronic Engineering, Tsinghua University, Beijing, China. He received the B.S. and M.S. degrees from Tsinghua University in 1986 and 1989, respectively, and the Ph.D. degree from The Hong Kong University of Science and Technology in 1998. His research interests include wireless communications and satellite communications. He is a Fellow of IEEE.

## References

[1] L. Kuang, J. Sun, J. Zhang, H. Cui, and K. Liu, "Towards Space-Based Computing Infrastructure Network: Development Trends, Network Architecture, Challenges Analysis, and Key Technologies." 2025. [Online]. Available: https://arxiv.org/abs/2503.06521

[2] C. Guimarães, A. Netti, M. Sauer, F. Zeiger, H.-P. Huth, and E. Boriskova, "A Survey on Satellite Computing: Connecting the Dots Between Networks and Applications," IEEE Commun. Surv. Tutor., vol. 28, pp. 567–592, 2026, doi: 10.1109/COMST.2025.3579525.

[3] Y. Zuo, M. Yue, H. Yang, L. Wu, and X. Yuan, "Integrating Communication, Sensing and Computing in Satellite Internet of Things: Challenges and Opportunities," IEEE Wirel. Commun., vol. 31, no. 3, pp. 332–338, 2024, doi: 10.1109/MWC.019.2200574.

[4] "Owl - Next-generation Monitoring | Planet." Accessed: Nov. 04, 2025. [Online]. Available: https://www.planet.com/constellations/owl/

[5] A. Lee, "How Starcloud Is Bringing Data Centers to Outer Space," NVIDIA Blog. Accessed: Jan. 11, 2026. [Online]. Available: https://blogs.nvidia.com/blog/starcloud/

[6] "Meet Project Suncatcher, a research moonshot to scale machine learning compute in space.," Google. Accessed: Jan. 11, 2026. [Online]. Available: https://blog.google/innovation-and-ai/technology/research/google-project-suncatcher/

[7] E. Berger, "Elon Musk on data centers in orbit: 'SpaceX will be doing this,'" Ars Technica. Accessed: Jan. 11, 2026. [Online]. Available: https://arstechnica.com/space/2025/10/elon-musk-on-data-centers-in-orbit-spacex-will-be-doing-this/

[8] "Beijing Institute to Build China's First Space Computing Center 800 km Above Earth." Accessed: Jan. 11, 2026. [Online]. Available: https://www.yicaiglobal.com/news/beijing-institute-to-build-chinas-first-space-computing-center-800-km-above-earth

[9] "Tsinghua's space network experimental platform achieves breakthrough-Tsinghua University." Accessed: Jan. 11, 2026. [Online]. Available: https://www.tsinghua.edu.cn/en/info/1399/14043.htm

[10] "Kubernetes," Kubernetes. Accessed: Oct. 31, 2025. [Online]. Available: https://kubernetes.io/

[11] "Amazon Elastic Kubernetes Service | AWS," Amazon Web Services, Inc. Accessed: Jan. 11, 2026. [Online]. Available: https://aws.amazon.com/cn/eks/

[12] "Tencent Kubernetes Engine | Tencent Cloud." Accessed: Jan. 11, 2026. [Online]. Available: https://www.tencentcloud.com

[13] "Container Engine Service CCE_Container Service_K8S_Container Management Platform-Baidu AI Cloud." Accessed: Jan. 11, 2026. [Online]. Available: https://intl.cloud.baidu.com/en/product/cce.html

[14] "KubeEdge Initiated by Huawei Cloud Becomes a CNCF Graduated Project," Huawei Cloud. Accessed: Jan. 11, 2026. [Online]. Available: https://www.huaweicloud.com/intl/en-us/news/20241018154136583.html

[15] Open Data Center Committee (ODCC), "China Comprehensive Computing Power Index (2024)." 2024. [Online]. Available: https://www.odcc.org.cn/news/p-1879463193510518785.html